\documentclass[doublecol]{epl2} 

\def\be{\begin{equation}}
\def\en{\end{equation}}

\def\ve{\varepsilon} 
\newcommand{\bi}[1]{\mbox{\boldmath$#1$}}

\def\bea{\begin{eqnarray}}
\def\ena{\end{eqnarray}}

\title{Nonionic 
and ionic surfactants at an interface}
\shorttitle{Surfactant adsorption} 

\author{Akira  Onuki}
\shortauthor{Akira  Onuki}

\institute{                    
Department of Physics, Kyoto University, 
Kyoto 606-8502, Japan
}

\abstract{
A Ginzburg-Landau 
 theory is presented on surfactants  in  
polar binary mixtures,  
which  aggregate at  an interface 
due to  the amphiphilic interaction. 
They can be ionic surfactants 
coexisting with counterions. 
Including the solvation and image interactions 
and accounting for a finite volume fraction 
of the surfactant,  
we obtain  their distributions and  
the electric potential around an interface in equilibrium. 
The surface tension is also calculated. 
The distribution of the adsorbed ionic surfactant 
is narrower than that of the counterions. 
The adsorption is 
marked for  hydrophilic 
and hydrophobic pairs of ionic  surfactant and counterions.  
 }

\begin{document}

\maketitle

\section{Introduction}

Surfactant molecules  
can be trapped at an interface,   
reducing the surface tension and 
 giving rise to mesoscopic 
structures, as their density  in the bulk 
is increased \cite{PG,Safran}. 
Oil-water 
interfaces containing surfactants are 
usually treated as infinitesimally thin 
surfaces  characterized by  the surface tension $\gamma$ 
dependent on the adsorbed surfactant density $\Gamma$. 
In this letter, we present a diffuse-interface model 
or a  Ginzburg-Landau  theory. 
Some authors 
presented  Ginzburg-Landau 
free energies  which include 
higher-order gradients of the  composition 
or the surfactant 
density to describe 
dynamics of mesophase formation  
\cite{Grant,Komura,Gompper,Yeomans}. 
However, the gradient 
expansion in the free energy breaks down  when the 
surfactants are strongly adsorbed 
onto an interface or when the trapping   
energy $\epsilon_{\rm a}$ of a surfactant 
 molecule much exceeds the 
thermal energy $T$. 
We will  propose a model in which 
surfactant molecules are  treated as rods and their 
 two ends can stay in  very different 
 environments.   In this letter,  we 
examine only the equilibrium properties 
following from our model.

Moreover,   surfactants can be  ions or have charges 
in many important systems in biology and technology.   
The  adsorption of surfactant and counterions 
in such cases has not yet been well studied.  
Hence we will  treat 
 ionic surfactants 
and  counterions in  binary mixtures, 
grafting  the amphiphilic interaction onto 
our previous theory of electrolytes 
which accounts for the solvation effect  
\cite{Onuki-Kitamura,OnukiPRE}.

We may define  the excess adsorption  
 $\Gamma$  generally for 
 doped particles  such as surfactants 
and ions.   For a  diffuse-interface,  we   determine 
the interface position in terms of the 
order parameter $\psi(z)$ (the composition difference in this work)  
  as  $ z_{\rm in}= 
\int_0^L dz [\psi(z)-\psi_\beta]/\Delta \psi$ 
 in  a  finite system 
in the region $0<z<L$, where all the quantities change 
along the $z$ axis \cite{Gibbs}.  
The  $\psi(z)$ takes the two bulk values 
$\psi_\alpha$ and $\psi_\beta$ with 
 $\Delta\psi\equiv  \psi_\alpha-\psi_\beta>0$ 
 on the two sides 
of the interface, while the bulk values of 
the doped-particle density $n(z)$    are denoted by 
$n_\alpha$ and $n_\beta$. The phase 
$\alpha$ is more polar or water-rich, while   
 the phase $\beta$ is less  polar or oil-rich. 
Then, 
\bea 
\Gamma &=& \int^{L}_{0}dz \bigg[n(z)-n_{\alpha} 
- \frac{\Delta n}{\Delta\psi}
(\psi(z)-\psi_\alpha)\bigg] \nonumber\\
&=&\int^{z_{\rm in}}_{0}dz [n(z)-n_{\alpha}] 
+ \int_{z_{\rm in}}^{L}dz [n_(z)-n_{\beta}], 
\ena   
where $\Delta n=n_\alpha-n_\beta$ and the integrands 
tend to zero far from the interface. 
At very small $n$,  
the surface tension decreases as  
$\gamma\cong \gamma_0-T \Gamma$,  
where $\gamma_0$ is the surface tension without 
doping \cite{Safran,Gibbs}.  For electrolyte systems  
\cite{OnukiPRE} there is also an electrostatic contribution,  
\be 
\gamma\cong  \gamma_0 -T\Gamma - \frac{1}{8\pi}
\int dz{\ve}|\nabla \Phi|^2,
\en 
where $\ve$ is the dielectric constant and 
$\Phi$ is the electric potential tending to 
constants far from the interface. 
We will   derive eq.~(2) including the amphiphilic interaction.
The Boltzmann constant will be set equal to unity.

\section{Nonionic 
surfactant in a binary mixture}

We first consider a polar binary 
 mixture  (water and oil) 
with a small amount of neutral surfactant. 
The water, oil,  and surfactant densities 
are $n_A$, $n_B$, and $n_1$, respectively. 
Assuming the same molecular 
size $a$ for water and oil, 
their  volume fractions 
are $\phi_A=a^3n_A$, $\phi_B=a^3n_B$, and 
$N_1 c_1= v_1 n_1$, respectively, where 
 $v_1$ is the  surfactant molecular volume. 
We define the normalized surfactant density 
$c_1=a^3n_1$.  Here the volume ratio  $N_1=v_1/a^3$ can be 
arbitrary.      We assume the space-filling condition 
$\phi_A+\phi_B+ N_1 c_1=1$; then, 
\be 
\phi_A= (1-N_1c_1)/2+\psi, \quad  
\phi_B= (1-N_1c_1)/2-\psi.
\en 
Here   $2\psi=\phi_A-\phi_B$ is  the 
composition difference of the solvent. 
We may set $\phi_A\cong  1/2+\psi$ and $\phi_B\cong  1/2 -\psi$ 
 neglecting  the surfactant volume fraction 
for $N_1c_1\ll 1/2-|\psi|$.  
The  free energy 
$F$ is the space integral of 
its density $f_{\rm non}$ of the form, 
\bea 
\frac{a^3}{T}f_{\rm non} 
&=&  \phi_A\ln \phi_A 
 + \phi_B\ln \phi_B  + \chi \phi_A\phi_B 
+ \frac{C}{2}|\nabla \psi|^2\nonumber\\   
&&+c_1 \ln c_1 - g_1 c_1\psi - c_1 \ln Z_{\rm am} ,  
\ena 
where $\chi$ is the temperature-dependent coefficient, 
 the coefficient $C$ is positive, and 
$g_1$ represents the 
interaction between the surfactant and the composition difference. 
The  last term represents the amphiphilic 
interaction, where $Z_{\rm am}$ is the partition function of 
 a rod-like dipole  with its center at the position $\bi r$. 
We assume that the surfactant molecules take a rod-like shape  
with length $2\ell$  considerably longer than $a$. 
It is given by the  following integral on  the surface of a sphere 
with radius $\ell$,    
\be 
Z_{\rm am}({\bi r}) = \int\frac{{d\Omega}}{4\pi} 
 \exp \bigg [- 
w_{\rm a} \psi({\bi r}+\ell {\bi u})+ 
w_{\rm a} \psi({\bi r}-\ell {\bi u})\bigg], 
\en 
where  $\bi u$  is the unit vector along the rod direction 
and $\int d\Omega$ represents the integration 
over the angles of $\bi u$. 
The  two ends of the rod  are at 
${\bi r} + \ell {\bi u}$ and 
${\bi r} -\ell {\bi u}$  under the influence of 
 the solvation potentials given by   
$T w_{\rm a} \psi({\bi r}+\ell {\bi u})$  
and 
$-T w_{\rm a} \psi({\bi r}-\ell {\bi u})$, respectively. 
It is instructive 
to examine the case in which 
 $\psi({\bi r})$ varies  slowly. That is, if the expansion  
 $\psi({\bi r}+\ell {\bi u})
- \psi({\bi r}-\ell {\bi u})\cong  2\ell {\bi u}
\cdot\nabla\psi$ is used, the last term in eq.~(4)  
becomes    
$ -\frac{2}{3}w_{\rm a}^2\ell^2|\nabla\psi|^2c_1,  
$. This gradient form 
was assumed in the literature \cite{Grant,Komura}. 
Together with the fourth term in eq.~(4), 
the coefficient in front of  $|\nabla\psi|^2$ 
vanishes for  $n_1=n_{L}$  (a Lifshitz  point), where  
\be 
n_{L}={3C}/[{4a^3(\ell w_{\rm a})^2}].  
\en 
For  $n_1>n_{L}$, 
a homogeneous solution is  
 unstable at a finite wave number, leading to a mesophase. 
If $C \sim a^{2}$, we have 
$a^3 n_{L}\sim (a/\ell w_{\rm a})^2$, so $n_{L}$ 
 is  small for  $\ell w_{\rm a} \gg a$.

The above  gradient expansion cannot be used 
around an interface far 
from the critical point or for $\xi <2\ell$. 
If a surfactant molecule is  
 trapped at an interface with  its 
hydrophilic (hydrophobic) 
 end in the water-rich (oil-rich)  region,  
the  free energy decreases by      
$\epsilon_{\rm a}=  Tw_{\rm a} \Delta\psi$. 
Strong adsorption occurs for large 
$\epsilon_{\rm a}/T=  w_{\rm a} \Delta\psi.$  
In  the one-dimensional (1D) case,  
where all the quantities vary along the $z$ axis, 
$Z_{\rm am}$ becomes 
\be 
Z_{\rm am}(z)=   
 \int_{-\ell}^\ell  \frac{ {d\zeta}}{2\ell } \exp [- 
w_{\rm a} \psi(z+\zeta)+ 
w_{\rm a} \psi(z-\zeta)], 
\en 
where  we have replaced 
$\psi({\bi r}\pm \ell {\bi u})$ in eq.~(5) by $\psi(z\pm \zeta)$ with 
$\zeta=\ell u_z$. In  the thin interface limit  $\xi\ll\ell$, 
we place  the interface at $z=0$  to  find    
\be 
 Z_{\rm am} (z) \cong 1+ ({1-|z|/\ell}) [\cosh({w_{\rm a} \Delta\psi})-1] 
\en 
for $|z|<\ell$, while  $Z_{\rm am} = 1$ 
for $|z|>\ell$.  Furthermore,  
in the dilute limit $N_1c_1\ll 1-\psi_\alpha$, 
we have $c_1(z)= c_{1\alpha}Z_{\rm am}(z)$ for $z<0$ 
and  $c_1(z)= c_{1\beta}Z_{\rm am}(z)$ for $z>0$, where 
$c_{1\alpha}$ and $c_{1\beta}=e^{-g_1\Delta\psi}c_{1\alpha}$  
are the bulk surfactant densities, leading to  
$ 
\Gamma\cong  \ell a^{-3}  (c_{1\alpha}+c_{1\beta}) 
[\cosh({w_{\rm a} \Delta\psi})-1]/2$. 
However, the steric effect due to finite 
surfactant volume becomes important with 
increasing $\Gamma$.

In this letter, we are interested 
in the equilibrium interface profile 
in the 1D case.   We require homogeneity 
of the surfactant chemical 
potential $T\nu_1= (\delta F/\delta n_1)_\psi$ 
and  the chemical potential 
difference $Th= a^3 (\delta F/\delta \psi)_{n_1}$ 
of the mixture. Some calculations give 
\bea 
{\nu_1}=&& \ln (c_1/Z_{\rm am} ) -g_1 \psi
-\frac{1}{2}N_1\ln (\phi_A\phi_B)\nonumber\\
&&+ 
\frac{1}{2}\chi N_1(N_1c_1+1)+
1-N_1,
\ena 
\be 
 h =\ln (\phi_A/\phi_B) - 2\chi \psi-C\nabla^2\psi- 
g_1 c_1 + h_{\rm am}.
\en 
The  ${h_{\rm am}}$ stems 
from the amphiphilic free energy 
$F_{\rm am}= -T \int d{\bi r} n_1\ln Z_{\rm am}$.  Its  
 1D form is   
\bea 
{h_{\rm am}(z)} =&& {w_{\rm a}} 
\int_{-\ell}^\ell \frac{ {d\zeta}}{2\ell } 
X (z+\zeta)\nonumber\\
&&\hspace{-2cm}\times
\bigg [  e^{w_{\rm a}[ \psi(z+2\zeta)-\psi(z)]}
- e^{w_{\rm a}[ \psi(z)-\psi(z+2\zeta)]}\bigg ],  
\ena 
where   $X(z)\equiv c_1(z)/Z_{\rm am}(z)$.  
In the dilute limit  $c_{1\alpha}\rightarrow 0$, 
$c_1(z)$ is expressed in terms of $\psi(z)$ as   
\be 
\frac{
c_1(z)}{c_{1\alpha}} \rightarrow \bigg [
\frac{1-4\psi(z)^2}{1-4\psi_\alpha^2}\bigg]^{N_1/2}
Z_{\rm am}(z)
e^{g_1(\psi(z)-\psi_\alpha)}. 
\en 
In our model, 
if $N_1$ is large and 
if $\psi_\alpha$ is close to $1/2$ far below the critical point, 
the first factor in the right hand side of eq.~(12) 
can be  large 
in the interface region where $\psi \cong 0$. This 
leads to $\Gamma \sim c_{1\alpha} 
\xi(1-4\psi_\alpha^2)^{-N_1/2}$ even for    
$Z_{\rm am}=1$.  For  our choice $\chi=3$ in our figures, 
we have $\psi_\alpha=0.429$ and 
$(1-4\psi_\alpha^2)^{-1}= 3.79$. 
However, the first factor in  eq.~(12)  tends to unity near 
the critical point.

The grand potential  $\Omega$ is  given by the 
space-integral of 
$\omega= f_{\rm non} - Ta^{-3}(h \psi+ \nu c_1)$. 
In equilibrium, $\Omega$  is minimized as a 
functional of $\psi$ and $c_1$ 
under given boundary conditions.  
For an interface, 
$\omega(z)$ should tend to a common limit $\omega_\infty$ 
as $z \rightarrow\pm \infty$ 
(see the appendix) and the surface tension  
 is  expressed as  
$\gamma  =\int dz [\omega(z)-\omega_\infty].$ 
Here it is convenient to 
introduce $\eta \equiv  
a^3 f_{\rm non}/T - \nu_1 c_1+c_1
= a^3 \omega/T + h \psi +c_1$. Use of eq.~(9) yields 
\bea 
\eta 
&=& (\frac{1}{2}+{\psi})\ln \phi_A +(\frac{1}{2}-{\psi})\ln \phi_B 
+N_1c_1\nonumber\\  
&&-\chi\psi^2-\frac{1}{4}\chi N_1^2c_1^2    +\frac{1}{2}C 
(\psi')^2 , 
\ena
where $\psi'=d\psi/dz$.  
The $\gamma$ can be expressed as 
\be 
\gamma= \frac{T}{a^{3}} \int dz \bigg[
\eta(z)- \eta_\alpha - h(\psi(z)-\psi_\alpha) 
-c_1(z)+c_{1\alpha} \bigg] ,
\en  
where 
$h= (\eta_\alpha-
\eta_\beta-c_{1\alpha}+c_{1\beta})/(\psi_\alpha-\psi_\beta)$ 
with $\eta_\alpha$, $\eta_\beta$, $c_{1\alpha}$, and 
$c_{1\beta}$ being the bulk values. 
The integrand in eq.~(14) 
tends to zero far from the interface. 
We may then  derive  eq.~(2) 
for small $c_1$.  
 Let $\psi_0(z)=\lim_{c_1\rightarrow 0}\psi(z)$ 
be the interface profile 
without surfactant 
and $\delta\psi=\psi-\psi_0$ be the deviation due to 
$c_1$. Then  $\eta(z) 
- \eta_\alpha = Cd/dz[\delta\psi d\psi_0/dz] +\cdots$ 
up to first order in $\delta\psi$ and $c_1$, because of   
the interface equation 
$\ln[(1+2\psi_0)/(1-2\psi_0)]-2\chi\psi_0 -C d^2\psi_0/dz^2 
=0$ without surfactant.   Hence 
 $a^3\Delta\gamma/T$ is nearly the integral of 
$-c_{1}(z)+c_{1\alpha}-  h (\psi-\psi_\alpha)$  
with  $h= (c_{1\beta}-c_{1\alpha})/\Delta\psi$, 
leading to eq.~(2). 
\begin{figure}
\onefigure{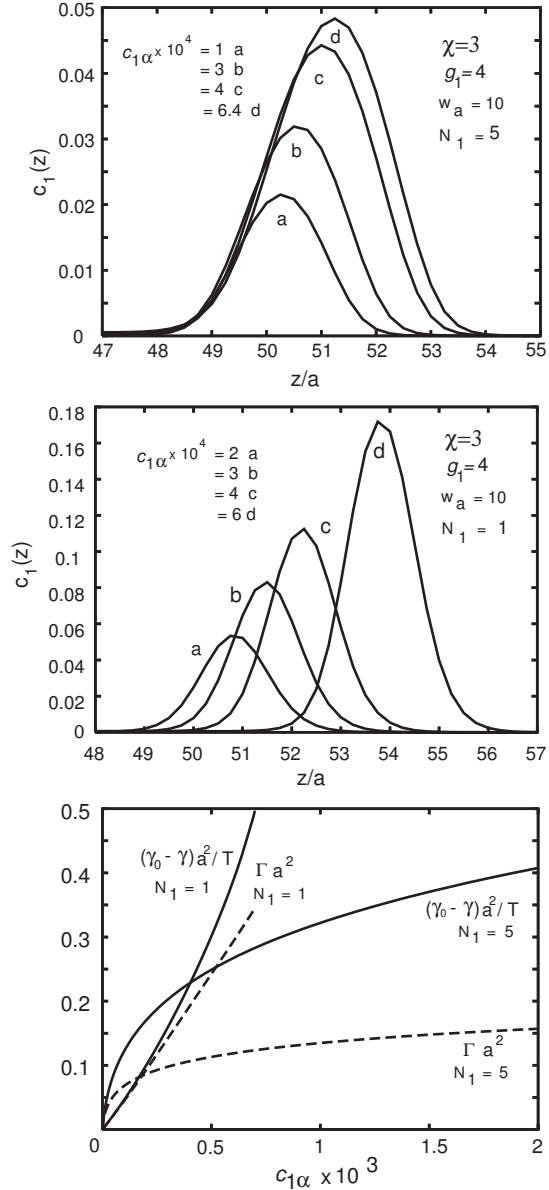}
\caption{Results for nonionic surfactants  for $\chi=3$, 
$g_1=4$, and $w_{\rm a}=10$: 
$c_1(z)$ for four values of 
$c_{1\alpha}$ for $N_1=5$ (top) and $1$ (middle), while 
$(\gamma_0-\gamma) a^2/T$ and $\Gamma a^2$  
versus $c_{1\alpha}$ for $N_1=5$  and 1 (bottom).  
Here $\gamma_0-\gamma \cong T\Gamma$ 
at small $c_{1\alpha}$.}
\label{fig.1}
\end{figure}

In all the figures to follow,  we will 
set $C=\chi a^2$, $\chi=3$, and  $\ell=2.5a$, where 
$\gamma_0= 0.497Ta^{-2}$ at $c_1=0$.
In eq.~(6) we have the critical density $n_L
= 0.36w_{\rm a}^{-2}$.  
The interface width $\xi$ is about $5a \sim 2\ell$.
In fig.~1, we present   $c_1(z)=a^{3}n_1(z)$,  
$(\gamma_0-\gamma) a^2/T$, and $\Gamma a^2$, where 
 $g_1=4$ and $w_{\rm a}=10$.  
With increasing $c_{1\alpha}$, 
 the peak of $c_1(z)$ grows,     $\Gamma$   increases,  
and $\gamma$ decreases. The steric effect 
due to  the surfactant volume fraction 
is apparent  for $N_1=5$, while it is not  
for $N_1=1$.  
For $N_1=5$, the first factor in eq.~(12) is crucial 
for $c_{1\alpha}<10^{-4}$ and 
the steric effect  is relevant 
for larger $c_{1\alpha}$, so that  eq.~(2) holds 
only for very small $c_{1\alpha}$;in fact,  
$(\gamma_0-\gamma)/T\Gamma 
\sim  1.3$  at $c_{1\alpha}=10^{-4}$. 
 See the argument below eq.~(12). 
For $N_1=1$,  eq.~(2)  nicely holds  
for $c_{1\alpha}<3\times 10^{-4}$ 
and $\Gamma\propto c_{1\alpha}$ at any $c_{1\alpha}$.

\section{Ionic surfactant and counterions 
 in a binary mixture}

Previously,  
we treated   non-amphiphilic ions,  
including the solvation 
and image interactions  \cite{Onuki-Kitamura,OnukiPRE}.  
 Here  the first ion species with density 
$n_1$ is a cationic surfactant. 
The  second species  with density 
$n_2$ constitutes 
anionic counterions having no amphiphilic character. 
They are both monovalent with charges $\pm e$.
We assume that the counterions are very  small and 
their volume fraction is negligible. Then  the 
relations in eq.~(3) still hold.

For our complex ionic system, 
 the total free energy density 
$f=f_{\rm non}+ f_{\rm ion}$ is the sum of 
$f_{\rm non}$ in eq.~(4) and the following new part, 
\be 
 f_{\rm ion}= 
Tn_2 (\ln c_2-g_2\psi)+ \frac{\ve(\psi)}{8\pi}{|\nabla \Phi|}^2 
 + n \mu_{im}, 
\en  
where $n=n_1+n_2$. We define $c_1=a^3n_1$ and 
$c_2=a^3n_2$.  The dielectric constant 
is  assumed to be of  the linear form 
  $\ve(\psi)=\ve_c+\ve_1\psi$, 
where $\ve_c$ and $\ve_1$ are positive 
constants.  The parameters $g_1$ in eq.~(4) 
and  $g_2$ in eq.~(15) can arise from the  solvation  (ion-dipole) 
interaction and can be very 
large in  aqueous solutions\cite{Onuki-Kitamura,OnukiPRE}. 
The differences $\Delta\mu_{\alpha\beta}^i= 
Tg_i \Delta\psi$  
are the so-called Gibbs transfer free energies 
per particle  from phase $\alpha$ to phase 
$\beta$ for ion species $i$ \cite{Hung}. 
For water-nitrobenzene mixtures at room temperatures, 
$\Delta\mu_{\alpha\beta}^i/T$ are of order 15  
for monovalent metallic ions. 
Recently,  
Sadakane {\it et al.}  
\cite{Seto}   found  periodic structures  
in  a near-critical  mixture containing 
strongly hydrophilic  and 
hydrophobic ions, presumably, 
with $g_1\sim -g_2 \sim 15$.

The electric potential $\Phi$ 
arises from the  charge density,  
\be 
-\nabla\cdot\ve(\psi)\nabla \Phi=
 4\pi e(n_1-n_2).  
\en 
The electric field $E= -d \Phi/dz$ around an interface reads   
\be 
 \frac{e}{T}E(z)  
 = \frac{4\pi \ell_{\rm B}}{1+{\ve_1}\psi(z)/{\ve_c}}
\int_{-\infty}^zd z'[n_1(z')-n_2(z')],
\en  
where the lower bound of the integration 
is pushed to $-\infty$.  We define the Bjerrum length 
$\ell_{{\rm B}}=e^2/\ve_c T$ 
 at $\ve=\ve_c$. We may set $\Phi_\alpha=0$   
without loss of generality.

The image potential  $\mu_{\rm im}$  
acts on  each ion
\cite{Onsager,Levin}, 
which arises  from inhomogeneous  $\ve$.  For moderate 
inhomogeneity and in the 1D case, 
it is expressed in the following Cauchy 
integral\cite{OnukiPRE},  
\be 
\mu_{\rm im}(z)= T Aa  
\frac{\ve_1 }{\ve_c} 
\int \frac{dz'}{\pi} 
\frac{{e^{-2\kappa |z-z'|}} }{z-z'} \frac{d\psi(z')}{dz'} ,
\en 
where  $A$ represents the charge stength, 
\be 
A=\pi e^2 /4a\ve_cT= \pi\ell_{\rm B}/4a, 
\en
and   $\kappa = 
[4\pi n e^2/\ve_{\rm c}T]^{1/2}$ is the Debye  
wave number \cite{Onsager}.
    The factor 
$e^{-2\kappa |z-z'|}$ in eq.~(18) 
arises from the screening 
of the image potential by the other ions, so the image 
interaction is weakened with increasing the ion density.
We  take  $n$ in $\kappa$ as the space-dependent ion density 
$n(z)=n_1(z)+n_2(z)$.

\begin{figure}
\onefigure{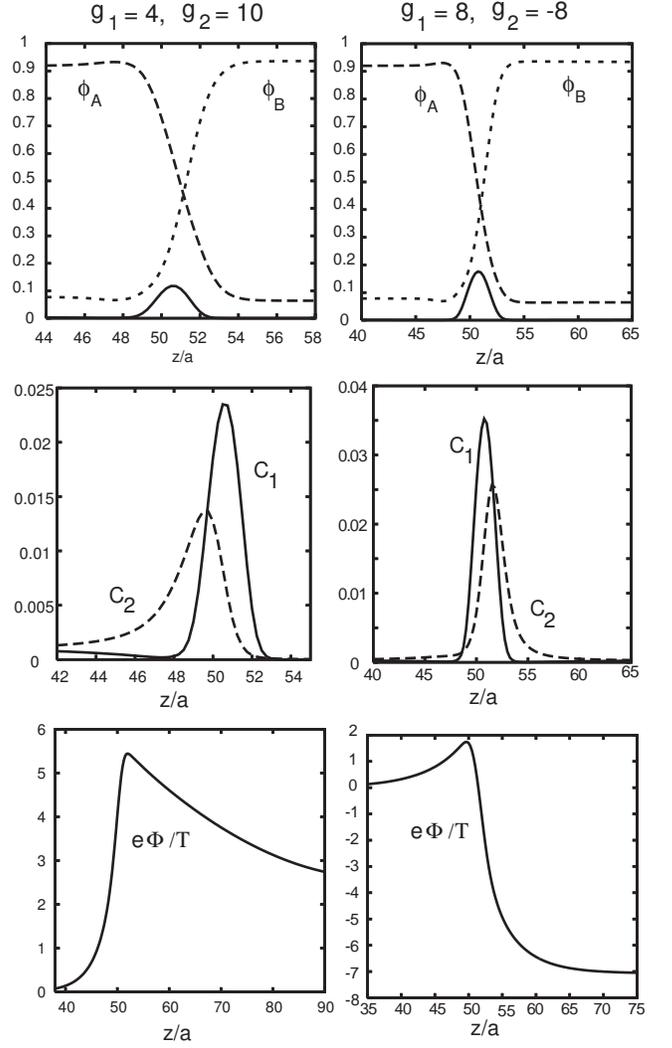}
\caption{ Profiles for mixtures with 
cationic surfactant and anionic counterions 
with  $N_1=5$, $w_{\rm a}=12$, and $A=4$. 
Top: $N_1c_1$ (bold line), $\phi_A$, and $\phi_B$. Middle:  
$c_1$ and $c_2$. Bottom: 
$e\Phi/T$.  Here  
$g_1=4$, $g_2=10$, and   $c_{1\alpha}=10^{-3}$ (left), 
while  $g_1=-g_2=8$ and  $c_{1\alpha}=3.6\times 10^{-4}$ (right). 
The counterion distribution 
has a peak in the phase $\alpha$ (left) 
or $\beta$ (right) depending on $g_2$. }
\label{fig.2}
\end{figure}

\begin{figure}
\onefigure{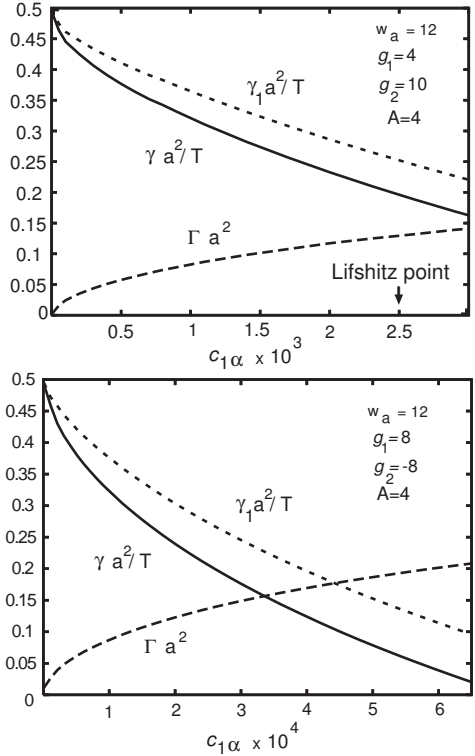}
\caption{
$\gamma a^2/T$, $\gamma_1 a^2/T$,  and   
$\Gamma a^2$ as functions of $c_{1\alpha}$ 
with  $N_1=5$, $\chi=3$,  and $w_{\rm a}=12$, where 
$\gamma$ is calculated from eq.~(26) and 
 $\gamma_1$ is the first term on its  right hand side.  
The curves change on a scale of 
$10^{-3}$ for $g_1=4$ and $g_2=10$ (top) 
and on a scale of 
$10^{-4}$ for $g_1=-g_2=8$ (bottom).}
\label{fig.3}
\end{figure}
\begin{figure}
\onefigure{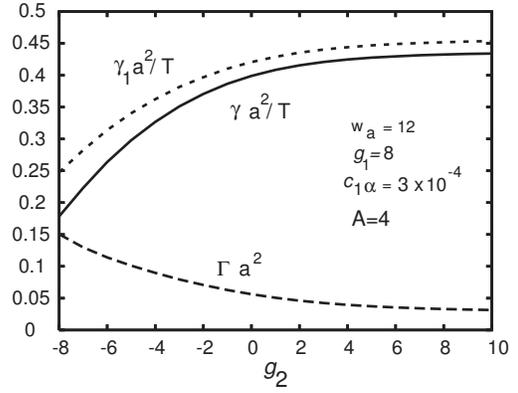}
\caption{
$\gamma a^2/T$, $\gamma_1 a^2/T$,  and   
$\Gamma a^2$ as functions of 
$g_2$ 
with  $\chi=3$, $w_{\rm a}=12$, and $g_1=8$,  where 
 $\gamma_1$ is the first term on the right hand side of eq.~(26).  
The counterions are hydrophobic for 
$g_2<0$ and  hydrophilic  for $g_2>0$.}
\end{figure}

Again we assume homogeneity of 
the chemical potentials $T\nu_{1{\rm tot}}= 
\delta F/\delta n_1$,   $T\nu_{2{\rm tot}} = 
\delta F/\delta n_2$,   and $Th_{\rm tot}= 
a^3\delta F/\delta \psi$. From 
$F=\int d{\bi r}(f_{\rm non}+ f_{\rm ion})$ they are     
\bea 
\nu_{1{\rm tot}} 
&=& \nu_1+ {e}\Phi/T +\mu_{\rm im}/T, \nonumber\\ 
\nu_{2{\rm tot}} &=&
\ln c_2+1-g_2 \psi - {e} \Phi/T  
+\mu_{\rm im}/T, \nonumber\\
h_{\rm tot} &=& h -g_2 c_2 + h_{\rm im}/T,
\ena 
where $\nu_1$ and $h$ are given by eqs.~(9) and (10). 
In taking the derivatives with respect to $n_1$ and $n_2$,  
we neglect the $n$-dependence of $\mu_{\rm im}$ in eq.~(18). 
The $h_{\rm im}(z)$ in eq.~(20) arises from the image interaction 
and is obtained by the right hand side of eq.~(18) if 
$\psi$ is replaced by 
$n= n_1+n_2$.

Since the counterion volume fraction 
is neglected, $c_2$  around an interface 
is written as 
\be 
\frac{c_2(z)}{ c_{2\alpha}}
=  \exp\bigg[g_2(\psi(z)-\psi_\alpha)+\frac{e}{T}\Phi(z)-
\frac{1}{T} \mu_{\rm im}(z)\bigg].   
\en 
In our previous work  \cite{OnukiPRE},
 we examined the image factor 
$F_{\rm ima}(z)\equiv 
\exp[ -\mu_{\rm im}(z)/T]$  in various cases. 
Let us consider it in the  $\alpha$  region with 
$z_{\rm in}-z >\xi$, where 
$z_{\rm in}$ is the interface position 
and $\xi$ is the interface thickness. Then, 
\be 
F_{\rm ima}(z) \cong 
\exp[-D
e^{-2\kappa_\alpha (z_{\rm in}-z)}/(z_{\rm in}-z)],
\en 
where  $D=\ell_{\rm B}\ve_1\Delta\psi/4\ve_c$. 
Here $D\sim  \ell_{\rm B}\sim Aa$ for 
$\ve_1 \sim \ve_c$ and $\Delta\psi \sim 1$. 
The image interaction can be crucial  
near an interface for  $\xi < D <  
\kappa_\alpha^{-1}$,  under which 
 the image  factor  serves to 
repel  the ions   in the region 
$\xi< z_{\rm in}-z < (2\kappa_\alpha)^{-1}$ in the 
$\alpha$ region.  This ion depletion 
was used to  explain  an increase of 
$\gamma$ of  water-air interfaces  
with  salt \cite{Onsager,Levin}. 
However, its efect on the ionic surfactants  
is diminished  when  the amphiphilic interaction 
is strong or for $w_{\rm a}\Delta \psi \gg 1$. 
These aspects will be illustrated 
in figs.~5 and 6 for $A=4$ and $10$.

From the charge neutrality in the bulk regions $\alpha$ 
and $\beta$, we require  $c_{1\alpha}=
c_{2\alpha}$ and $c_{1\beta}= c_{2\beta}$.  Then 
we find   
\be 
\ln ({c_{1\alpha}}/{c_{1\beta}})  
= (g_1+g_2)\Delta\psi/2+  S_t,
\en
\be 
e(\Phi_\alpha- \Phi_\beta)/T = 
(g_1-g_2)\Delta\psi/2+S_t, 
\en  
where  $\Phi_\alpha-\Phi_\beta$ 
is called the  Galvani potential difference.  
The $S_t$ stems  from the surfactant  volume 
fraction $N_1c_1$,   
\be 
S_t= \frac{N_1}{4}\ln \bigg[
\frac{(1-N_1c_{1\alpha})^2-
4\psi_\alpha^2}{(1-N_1c_{1\beta})^2-4\psi_\beta^2}\bigg] 
-\frac{N_1^2}{4}\chi
\Delta c_1, 
\en
 with $\Delta c_1=c_{1\alpha}-c_{1\beta}$. 
Here $S_t\rightarrow 0$ as $N_1\rightarrow 0$. 
Note that  $\Phi(z)$  
changes on the scale of  the Debye 
screening length,  $\kappa_\alpha^{-1}$ in the  phase 
$\alpha$  and $\kappa_\beta^{-1}$ in the   phase $\beta$. 
As a result,  $\Phi$ changes from 
$\Phi_\alpha$ to $\Phi_\beta$ on  the spatial scale  
 of $\kappa_\alpha^{-1}+\kappa_\beta^{-1}$.

We consider the grand potential density 
$\omega= f- Ta^{-3}(h_{\rm tot}\psi+
\nu_{1{\rm tot}}c_1+\nu_{2{\rm tot}} c_2)$. 
With the aid of eq.~(20) we have  
 $\omega= Ta^{-3} (\eta -  h_{\rm tot}\psi-c) 
+{\ve}|\nabla\Phi|^2/8\pi - \rho \Phi$, 
where  $\eta$ is given by eq.~(13)  and $c=c_1+c_2$. 
From $\omega(\infty)=\omega(-\infty)$ 
(see the appendix), we obtain  
\bea 
\gamma &=& \frac{T}{a^{3}} \int dz \bigg[
\eta(z) - h_{\rm tot}\psi(z) 
-c(z)-A_{\alpha} \bigg]\nonumber\\ 
&&- \frac{1}{8\pi}
\int dz {\ve(\psi)} |\nabla\Phi|^2,
\ena  
where   $A_\alpha= \eta_\alpha - h_{\rm tot}\psi_\alpha 
-c_\alpha$ and 
$h_{\rm tot}= (\eta_\alpha-\eta_\beta-c_\alpha+c_\beta)/\Delta\psi$ 
with $c_\alpha$ and $c_\beta$ being the bulk values of $c$. 
From this expression,  eq.~(2) 
follows at small  $n=n_1+n_2= a^{-3}c$ if we use the argument below 
eq.~(14).

We  present  some numerical results with  
 $N_1=5$ and $\ve_1=0.8\ve_c$. 
The dielectric constant of the phase $\alpha$ 
is twice  larger than that of the phase $\beta$ at $\chi=3$. 
We  set $A=4$ except in the left panel of fig.~6 
where $A=10$.\\  
(i) In fig.~2, we show the volume fractions  $\phi_A$, $\phi_B$, 
and $N_1 c_1=1-\phi_A-\phi_B$ (top),  
$c_1$ and $c_2$ (middle), 
and $e\Phi/T$ with $\Phi_\alpha=0$ 
(bottom).   In the  left plates, the counterions are 
more hydrophilic than the cationic surfactant, where 
 $g_1=4$ and  $g_2=10$ leading to 
  $\Gamma=0.124 a^{-2}$ and $\gamma   =0.317 Ta^{-2}$ at 
$c_{1\alpha}= 10^{-3}$. 
In the  right  plates, the  surfactant cations are 
hydrophilic and the counterions 
are hydrophobic, where      $g_1=-g_2=8$ leading to 
$\Gamma= 0.155 a^{-2}$ 
and $\gamma  = 0.159 Ta^{-2}$ 
 at $c_{1\alpha}=3.6\times 10^{-4}$.  
The distribution of the ionic  surfactant $c_1$ 
is narrower than that of the counterions $c_2$. 
This gives rise  to a peak of   $\Phi$, 
at which the right hand side of  eq.~(17) vanishes.\\  
(ii) In fig.~3, we show $\gamma$, $\gamma_1$, and $\Gamma$ 
as functions of $c_{1\alpha}$ at $w_{\rm a}=12$, where 
$\gamma_1$ is the first term on 
the right hand side of eq.~(26).  
The Lifshits point $c_{1\alpha}=a^3n_L=2.5\times 10^{-3}$ 
for $C=9a^2$  is 
marked by an arrow (top).  For small $c_{1\alpha}$ 
 eq.~(2) holds, but 
the last electrostatic term  
 in eq.~(26) is not negligible.  For a pair of  
hydrophilic  and hydrophobic ions, 
a large electric double layer is formed 
at an  interface\cite{OnukiPRE},  leading to  
a   large $\Gamma$ 
even at small $c_{1\alpha}$.\\ 
(iii) In fig.~4,  we show 
$\gamma$, $\gamma_1$, and $\Gamma$ 
versus  $g_2$ at $w_{\rm a}=12$ and $g_1=8$ 
to demonstrate the above  trend, where 
$\Gamma$ and $\gamma_0-\gamma$  
 are markedly  enhanced  for negative $g_2$.\\  
(iv)   
In figs.~5 and  6, we set  $w_{\rm a}=15$, where $\Gamma$ 
is appreciable even at   $c_{1\alpha}=10^{-5}$.     
In fig.~5, where   $g_1=g_2=10$,  
we have    
$\Gamma= 0.0090 a^{-2}$ and  $\gamma=0.487T/a^2$, while    
$\gamma_1=0.491T/a^2\cong \gamma_0-T\Gamma$. 
There are virtually no ions in the region $\beta$. 
The distribution of $c_2$ in the region $\alpha$ 
is  broad  changing on the scale of 
$\kappa_\alpha^{-1}$, where   
$\kappa_\alpha= 0.046/a$. 
Note  that $\Phi(z)$ is  constant 
in the region $\beta$ displayed  in fig.~5. 
However, it should tend  to $\Phi_\beta(\cong 0$ here)  
for   $z-z_{\rm in} \gg 
\kappa_\beta^{-1} \sim 10^3a$.\\ 
(v) In the left panel of fig.~6, 
we increase $A$ in eq.(19) to 10 with the other 
parameters being the same as in fig.~5. 
This is because  the effect of the  image interaction 
is rather weak at $A=4$\cite{OnukiPRE}, where $D$ in eq.~(22) 
is of order $a$.
 See the discussion around eq.~(22).  
With $A=10$, the image interaction  is  amplified  
and $\Gamma$ is decreased to $0.0049a^{-2}$ 
($55\%$ of the value in fig.~5) with  
$\gamma=0.491T/a^2$. See fig.~10 of our previous 
paper\cite{OnukiPRE}, where we set $A=10$ 
to realize strong ion depletion 
in the more polar phase. However,  the normalized potential 
$e\Phi/T$ is also amplified with increasing $A$ from eq.~(17) 
and the effect is very complicated. 
 \\
(vi)  
In the right panel of fig.~6, 
we consider the case of 
a hydrophilic and hydrophobic ion pair  with 
 $g_1=-g_2=8$. The  other 
parameters are the same as in fig.~5. 
Then $\Gamma$ is  incresaed 
to $0.071a^{-2}$ ($8$ times larger than 
 the value in fig.~5) 
with $\gamma= 0.403T/a^2$.

\begin{figure}
\onefigure{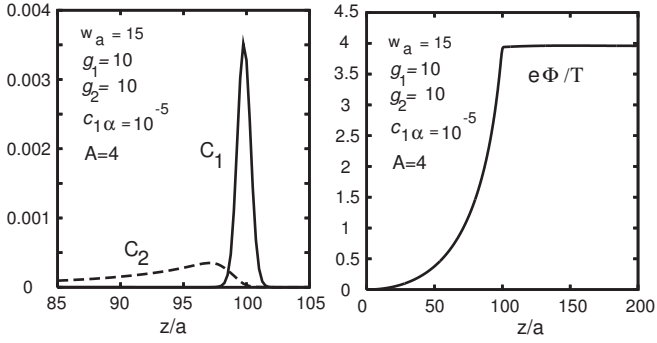}
\caption{$c_1$ and $c_2$ (left) and 
$e\Phi/T$ (right) with   
$w_{\rm a}=15$,  
$A=4$,  and $g_1=g_2=10$,   
where $c_{1\alpha}=10^{-5}$ is very small due to large $w_{\rm a}$. 
The  counterion distribution is 
  wider than that of the ionic surfactant. 
There is no appreciable variation of  $\Phi$ 
in the $\beta$ region shown here, where the ion densities 
are nearly zero.}
\end{figure}      

\begin{figure}
\onefigure{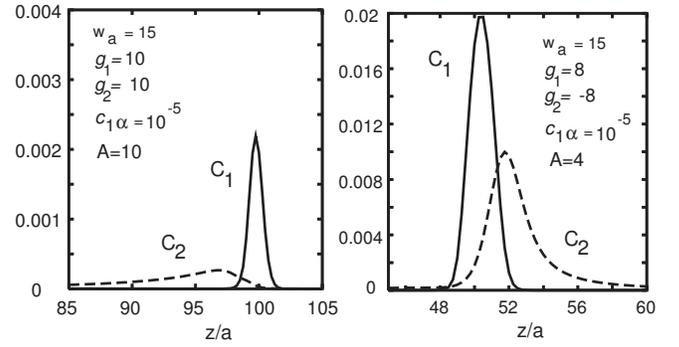}
\caption{
$c_1$ and $c_2$ 
for $w_{\rm a}=15$ and 
$c_{1\alpha}=10^{-5}$. Left:  
 $A=10$, where the electrostatic and 
image interactions are amplified from eqs.~(17) and (18). 
 Right:  $g_1=-g_2=8$ leading to enhanced adsorption. 
The other parameters in each panel are 
the same as in fig.~5. 
}
\end{figure}

\section{Summary} 
We have presented  
the continuum models of surfactants including 
 the  amphiphilic interaction 
explained around eqs.~(4)-(8). They 
 reasonably describe the adsorption of nonionic and ionic 
surfactants, though our  numerical examples are still   
fragmentary.  For ionic surfactants, 
included also are the electrostatic, 
solvation,  and image 
interactions.  The adsorption 
is extremely sensitive to 
$\epsilon_a/T=w_{\rm a}\Delta\psi$ and $N_1$. 
It is decreased  as $A \sim \ell_{\rm B}/a$ 
is increased for  a pair of hydrophilic ions 
from  the left panels of figs.~5 and 6. 
It is also enhanced  for 
hydrophilic and hydrophobic ion pairs, 
 as shown in figs.~3 and 4 and the right panel of fig.~6.

Though we have set 
$\chi=3$ in all the figures here, 
the changeover and the phase transition 
with decreasing  $\chi$ 
at large $w_{\rm a}$  
and $g_i$   should be     
intriguing \cite{OnukiPRE}. 
The phase transition of binary mixtures  
with salt can be  very complex  
at large $g_i$ 
\cite{Onuki-Kitamura,OnukiPRE,Seto}.  
We  should also study 
dynamics including the amphiphilic interaction 
in higher space dimensions. 

\section{Appendix}
\setcounter{equation}{0}
\renewcommand{\theequation}{A.\arabic{equation}}

We consider a  1D eqilibrium interface  profile, 
 $\psi=\psi(z)$, $c_1= c_1(z)$, 
and $c_2= c_2(z)$,  for our ionic surfactant system. 
Use of  eqs.~(9), (10), and (18) gives
\bea 
&&\frac{d}{dz}\bigg [\frac{a^3}{T} \omega  -{C} (\psi')^2+ 
(c_1-c_2) \frac{e\Phi}{T}  -\frac{c}{T}  \mu_{\rm im} \bigg]\nonumber\\
&&= - {h_{\rm int}}\psi' -\nu_{1{\rm int}}c_1'- \nu_{2{\rm int}}c_2', 
\ena 
where  $c=c_1+c_2$, 
$\psi'=d\psi/dz$, $c_{K}'= dc_K/dz$, and 
\be 
h_{\rm int}=\frac{a^3}{T}
\frac{\delta F_{\rm int}}{\delta\psi}, \quad 
\nu_{K{\rm int}} =\frac{a^3}{T}
\frac{\delta F_{\rm int}}{\delta c_K},
\en
with  $K=1,2$.  
Here $F_{\rm int}= \int d{\bi r}
[-T n_1\ln Z_{\rm am}+n\mu_{\rm im}]$ 
 is the sum of the amphiphilic and image  
free energies, so   
$
h_{\rm int}=
h_{\rm am}+h_{\rm im}$ and 
$\nu_{1{\rm int}}+
 \ln Z_{\rm am}=\nu_{2{\rm int}}\cong \mu_{\rm im}$ 
(see  the discussion below eq.~(20)). 
The  $F_{\rm int}$ is 
 invariant  with respect to a shift of the interface position 
or  with respect to $z\rightarrow z- \zeta$ in $\psi$ and $c_K$. 
Then 
$
\int dz [{h_{\rm int}}\psi'+  \nu_{1{\rm int}}c_1'
+\nu_{2{\rm int}}c_2']=0 
$  and integration of eq.~(A.1)  gives 
  $\omega(\infty)=\omega(-\infty)$.

\acknowledgments
This work was  
 supported by   Grants in Aid for Scientific 
Research and for  
the 21st Century COE project (Center for Diversity and Universality in
Physics) from the Ministry of Education, Culture, Sports, Science and 
Technology of Japan.


\begin{thebibliography}{0}


\bibitem{PG} 
\Name{de Gennes P. G.  and  Taupin C.}  
\REVIEW{  J. Phys. Chem. }{86}{1982}{ 2294}. 


\bibitem{Safran} 
\Name{Safran S. A.} 
\Book{\it Statistical Thermodynamics of Surfaces,   
Interfaces, and Membranes} 
\Publ{Westview Press}
\Year{ 2003}.  



\bibitem{Grant} 
\Name{Laradji M., Guo H., Grant M.,  and 
Zukermann  M.}  
\REVIEW{J. Phys. A}{ 24}{1991}{L629}. 



\bibitem{Komura} 
\Name{Komura S. and 
 Kodama H.}  
\REVIEW{Phys. Rev. E}{ 55}{1997}{1722}. 

\bibitem{Gompper} 
\Name{Theissen O., Gompper G, and Kroll D.M. }
\REVIEW{Europhys. Lett.}{42}{1998}{419} 

\bibitem{Yeomans} 
\Name{Lamura A., Gonnella G, and Yeomans J.M. }
\REVIEW{Europhys. Lett.}{45}{1999}{314} 


\bibitem{Onuki-Kitamura} 
\Name{Onuki A.  and  Kitamura H.}  
\REVIEW{  J. Chem. Phys.}{121}{2004}{ 3143}.

\bibitem{OnukiPRE} 
\Name{ Onuki A.} 
\REVIEW{Phys. Rev. E}{73}{2006}{ 021506}. 


\bibitem{Gibbs} \Name{Gibbs J.W.} 
\Book{Collected works, 
 vol.1}
\Publ{Yale University Press, New Haven, CT} 
\Year{1957} 
\Page{219-331}. 





\bibitem{Hung}  
\Name{Le Quoc Hung} 
\REVIEW{ J. Electroanal. Chem.}{ 115}{1980}{159}.


\bibitem{Seto} 
\Name{ Sadakane K., Seto H.,  Endo H., 
 \and Shibayama M.}  
\REVIEW{J. Phys. Soc. Jpn.}{ 76}{2007}{113602}. 



\bibitem{Onsager} 
\Name{ Onsager L.  and  Samaras N. N. T.} 
\REVIEW{J. Chem. Phys.}{ 2}{1934}{ 528}.

\bibitem{Levin}  
\Name{ Levin Y.  and  Flores-Mena J.E.}  
\REVIEW{Europhys. Lett.}{ 56}{2001}{ 187}.  

\end{thebibliography}
\end{document}